\documentclass[9pt,twocolumn,twoside]{pnas-new4arxiv}

\templatetype{pnasresearcharticle} 

\usepackage{graphicx}               
\usepackage[noabbrev]{cleveref}

\usepackage{xspace}
\newcommand{\kindex}[2]{\ensuremath{{#1}_{\scalebox{0.5}{#2}}}}
\newcommand{\Ae}{\textit{Ae}\xspace}

\newcommand{\supvid}[1]{supporting video~S{#1}}

\usepackage{siunitx}
\DeclareSIUnit\px{px}

\graphicspath{{../figurematter/latex/figs/}{../figurematter/plots/}}

\begin{document}

\title{Highly deformable flapping membrane wings suppress the leading edge vortex in hover to perform better}


\author[a]{Alexander Gehrke}
\author[a,1]{Karen Mulleners}

\affil[a]{École polytechnique fédérale de Lausanne, Institute of Mechanical Engineering, Unsteady Flow Diagnostics Laboratory, 1015 Lausanne, Switzerland}

\leadauthor{Gehrke} 

\significancestatement{
When insects flap their wings, they typically create a vortex on top of the wings which helps to keep them aloft. 
When bats flap their wings, the wings deform much more than insect wings.
Our study provides new insights on the role of the vortex when flapping wing are highly deformable. 
We have used flexible membrane wings that can create more lift than stiff wings by bending and changing shape when they are flapped back and forth. 
When these flexible wings bend enough but not too much, the flow follows the wing curvature without creating a vortex which increases the lift. 
Our findings help explain how bats fly efficiently and provide new ideas for designing and controlling human-engineered flying vehicles.
}

\authorcontributions{
\textbf{AG:} Conceptualisation, Investigation, Formal analysis, Writing - original draft, Visualisation;
\textbf{KM:} Funding acquisition, Supervision, Writing - Review \& Editing, Visualisation.\\
}
\authordeclaration{The authors declare that they have no conflict of interest.}
\correspondingauthor{\textsuperscript{1}To whom correspondence should be addressed. E-mail: karen.mulleners@epfl.ch}

\keywords{fluid-structure interaction $|$ membrane wings $|$ flapping wing flight $|$ leading edge vortex $|$ unsteady fluid dynamics}

\begin{abstract}
Airborne insects generate a leading edge vortex when they flap their wings. 
This coherent vortex is a low pressure region that enhances the lift of flapping wings compared to fixed wings.
Insect wings are thin membranes strengthened by a system of veins that does not allow large wing deformations. 
Bat wings are thin compliant skin membranes stretched between their limbs, hand, and body that show larger deformations during flapping wing flight. 
This study examines the role of the leading edge vortex on highly deformable membrane wings that passively change shape under fluid dynamic loading maintaining a positive camber throughout the hover cycle.
Our experiments reveal that unsteady wing deformations suppress the formation of a coherent leading edge vortex as flexibility increases.
At lift and energy optimal aeroelastic conditions, there is no more leading edge vortex.
Instead, vorticity accumulates in a bound shear layer covering the wing's upper surface from the leading to the trailing edge. 
Despite the absence of a leading edge vortex, the optimal deformable membrane wings demonstrate enhanced lift and energy efficiency compared to their rigid counterparts.
It is possible that small bats rely on this mechanism for efficient hovering. 
We relate the force production on the wings with their deformation through scaling analyses.
Additionally, we identify the geometric angles at the leading and trailing edges as observable indicators of the flow state and use them to map out the transitions of the flow topology and their aerodynamic performance for a wide range of aeroelastic conditions.
\end{abstract}


\maketitle
\thispagestyle{firststyle}
\ifthenelse{\boolean{shortarticle}}{\ifthenelse{\boolean{singlecolumn}}{\abscontentformatted}{\abscontent}}{}


\dropcap{I}n 1934, the French entomologist Antoine Magnan wrote in his book on insect flight that bumblebees should not be able to fly \cite{magnan1934locomotion}.
According to conventional fixed-wing aerodynamic theory, the bumblebee's small wings could not produce enough lift for sustained flight.
It took almost until the end of the last century and the invention of modern high-speed camera technology to uncover what allowed bumblebees and many other airborne insects to fly proficiently: the leading edge vortex~\cite{Ellington.1984, Dickinson.1993}.
Above a critical Reynolds number, any thin airfoil or flat plate at sufficiently high angles of attack is subject to flow separation over the wing's leading edge due to the strong streamwise pressure gradient and high shear at the sharp leading edge.
The shear layer separating from the leading edge is not convected downstream in a straight line, but rolls up into a vortex on top of the wing.
This leading edge vortex grows in size and strength when it is fed by a constant or increasing shear rate.
The pressure in the vortex core region is lower than the surrounding pressure and as long as the vortex is attached to the wing, it will augment the lift produced by flapping wings compared to fixed wings, resolving Magnan's conundrum \cite{Wang.2005, Bomphrey.2006, Ford.2013, eldredge_leading-edge_2019}.
Eventually, the leading edge vortex will separate from the wing and shed into the wake where it loses coherence and breaks down.
When the vortex leaves the wing, so does the lift augmentation associated with it. 
This marks the time for flapping wing fliers to reverse the direction of their wing motion and recreate another leading edge vortex on the other side of the wing that is now facing upward. 
By flapping their wings back and forth in addition to rotating them, many flapping wing fliers seem to ensure the presence of a lift enhancing vortex for extended parts of the flapping cycle~\cite{ellington_leading-edge_1996, Dickinson.1999, sane_aerodynamics_2003}.
However, most of the past research on the role of leading edge vortices for flapping wing flight is limited to rigid wings~\cite{eldredge_leading-edge_2019, liu_vortices_2024} even though natural fliers predominantly use compliant and deformable wings \cite{wootton_functional_1992}.

The wings of birds and bats are often compliant and highly articulated, which is associated with several advantages, including weight reduction, efficient propulsion, resilience to disturbances like gusts, and silent operation \cite{rayner_bounding_1985, swartz_wing_1992, lentink_how_2007, young_details_2009, Ramananarivo.2011, boerma_wings_2019, jaworski_aeroacoustics_2020}.
This makes highly deformable wings an attractive model for improving the design of micro air vehicles \cite{shyy_flapping_1999, abdulrahim_flight_2005, pines_challenges_2006, stanford_fixed_2008, shyy_recent_2010, bleischwitz_aspect-ratio_2015, ramezani_biomimetic_2017, bomphrey_insect_2018} and other aerodynamic applications like flapping wing energy harvesters \cite{mathai_fluidstructure_2022}.

Deformable membrane wings offer enhanced aerodynamic performance when compared to rigid wings due to their ability to passively camber and reduce their effective angle of attack \cite{song_aeromechanics_2008, jaworski_high-order_2012}.
Insect wings are thin membranes which are strengthened by a system of veins that does not allow large wing deformations. 
Flying and gliding mammals, such as bats, lemurs, and squirrels, have wings that consist of compliant skin membranes that are stretched between their limbs, hand, and body \cite{cheney_bats_2022}. 
Bats also possess a lightweight bone and muscle structure that allows for precise control and large shape deformations of their membranes wings, thereby regulating wing curvature to improve flight efficiency \cite{tian_direct_2006, wolf.2010, riskin.2010, bahlman_design_2013, muijres_leading_2014, hedenstrom_bat_2015, swartz_advances_2015}.
Based on respirometry data, nectar-feeding bats exhibit substantially lower metabolic costs — about 40\% less than hawkmoths and 60\% less than hummingbirds of similar size \cite{winter_energetic_1998, winter_energy_1998, voigt_energetic_1999}.
This suggests that bats may use even more efficient mechanisms for lift generation in hover than insects and birds.
The exact reasons behind the greater energetic efficiency of bat hovering, compared to equivalently sized insects and birds, remain unclear.
We hypothesize that the large deformations of the membrane wings of bats are key to their aerodynamic performance. 

Here, we experimentally demonstrate the potential of highly deformable membrane wing to enhance the aerodynamic performance of flapping wings in hover. 
We specifically focus on identifying changes in the leading edge vortex topology when wing compliance is increased and on relating the changes in the membrane deformation to the flow topology and aerodynamic forces.


\section*{Flexible membrane wing platform}

\begin{figure*}[tb]
\centering
\includegraphics[width=\textwidth]{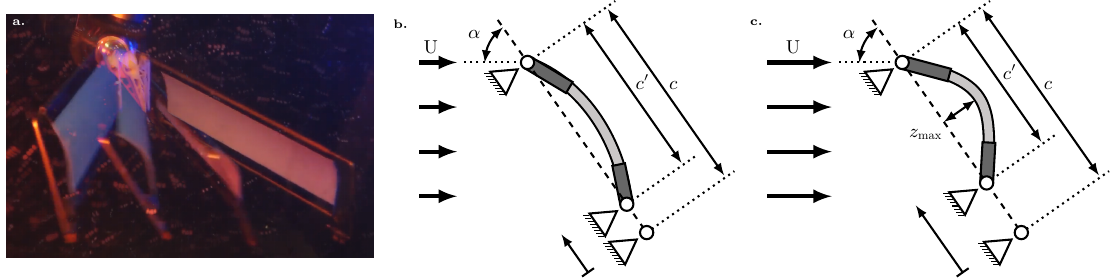}
\caption{
a.~Simulated multiexposure photo of the flapping flexible membrane wing. 
b. and c.~Two-dimensional representation of the passively deforming membrane wing for lower and higher flapping frequencies. Here, $c'$ represents the shortening of the chord length, and \kindex{z}{max} the maximum chord-wise camber due to an increase in the dynamic pressure ($q = 0.5 \rho U^2$).
}
\label{fig:flexWingIntroduction}
\end{figure*}%

In previous work \cite{gehrke_aeroelastic_2022}, we introduced a novel membrane wing platform to study unsteady aeroelastic interaction for flexible flapping wings.
The wing moves back and forth on a horizontal stroke plane similar to the wing kinematics of many insect, bird, and bat species in hovering flight (\cref{fig:flexWingIntroduction}), where the stroke and pitch angle kinematics are symmetric and the front- and backstroke are identical.
At the beginning of every stroke, the wing accelerates and creates a relative flow velocity parallel to the stroke direction of the wing (\cref{fig:flexWingIntroduction}b).
With increasing velocity, the pressure on the compliant membrane increases and the wing takes a curved shape, characterised by the chord-wise maximum camber \kindex{z}{max}, and shortens its effective chord length to $c'$ (\cref{fig:flexWingIntroduction}b-c).
In the second half of a stroke, the wing decelerates which leads to lower incoming dynamic pressure and the wing eventually returns to its undeformed, flat state.
Between the front- and back-stroke, the wing rotates around its pitch axis to maintain a positive geometric angle of attack $\alpha$.
(See \supvid{2 -S4} for videos of the flapping membrane wing in action.)

The aeroelastic response during the flapping cycle varies as a function of various parameters, including the wing's angle of attack $\alpha$, the membrane compliance ($E h$, with $E$ the Young's modulus and $h$ the thickness of the compliant membrane), and the stroke average wing velocity ($\bar{U} = 2f \hat{\phi} \kindex{R}{2}$, with $f$ the flapping frequency, $\hat{\phi}$ the peak-to-peak stroke amplitude, and \kindex{R}{2} the location of the second moment of area as the radial reference location.)
Bars on top of variables, $(\bar{\phantom{u}})$, indicate stroke-average quantities. 
Hats on top of variables, $(\hat{\phantom{u}})$, indicate cycle-maximum or -minimum quantities which are also referred to as amplitudes.

The ratio between the bending resistance of the membrane and the aerodynamic loading is characterised by the aeroelastic number for a wide range of material properties and wing motions \cite{waldman_camber_2017, gehrke_aeroelastic_2022}.
The aeroelastic number \Ae is defined as:
\begin{equation}
\Ae = \frac{E h}{\frac{1}{2} \rho \bar{U}^2 c} = \frac{E h}{2 \rho f^2 \hat{\phi}^2 \kindex{R}{2}^2 c} \, ,
\label{eq:aeroelastic_number}
\end{equation}%
where $c$ is the wing's chord-length, and $\rho$ the fluid density.
The Cauchy number ($\textit{Ca}$) and effective stiffness ($\kindex{\Pi}{1}$) are often used interchangeably in the literature for similar problems and directly relate to the aeroelastic number as $\Ae = \textit{Ca}^{-1} = \kindex{\Pi}{1}^3$.

A low aeroelastic number corresponds to a compliant wing or high aerodynamic loading relative to the bending resistance, and vice versa.
We estimate typical values of the aeroelastic number of bat wings based on data from three bat species that predominantly use hovering flight in their foraging activities: the Pallas's long-tongued bat (\textit{Glossophaga soricina}), the lesser long-nosed bat (\textit{Leptonycteris yerbabuenae}), and the Geoffroy's tailless bat (\textit{Anoura geoffroyi}) \cite{winter_operational_2003, cole_leptonycteris_2006, ortega_anoura_2008}. 
The estimates lie between $\Ae = 0.10$ and $4.33$, with an average around $\Ae = 1.67$ \cite{swartz_mechanical_1996, norberg_cost_1993, norberg_wing_2006, norberg_ecological_1987, sazima_two_2022}.
The range of aeroelastic numbers we tested here generously covers the values estimated for the bat species highlighted.
More details about the \Ae estimation and the data sources used and procedure followed are presented in the supporting information.

The Reynolds number \textit{Re} quantifies the ratio of inertial to viscous force contributions, and serves as an indicator for the development of flow structures across varying length and time scales.
In the context of hovering flight, \textit{Re} can be determined using the stroke average wing velocity at the radius of the second moment of area:
\begin{equation}
\textit{Re} = \frac{\bar{U} c}{\nu} = \frac{2 f \hat{\phi} \kindex{R}{2} c}{\nu} \, ,
\label{eq:reynolds_number}
\end{equation}%
where $\nu$ is the kinematic viscosity of the fluid.
The Reynolds number varies between \textit{Re} = \numrange[range-phrase = { and }]{2800}{9300} for the range of flapping frequencies tested in this study.
Our experiments cover the typical Reynolds number range for hovering of larger flying insects such as the hawkmoth (\textit{Manduca sexta}), small avian species like the rufous hummingbird (\textit{Selasphorus rufus}), and small bat species, including the aforementioned Pallas's long-tongued bat \cite{shyy_recent_2010}.
In this Reynolds number regime, laminar flow separation and coherent vortex formation occur for flat plates at high angles of attack and fixed wing aircrafts produce less lift compared to flapping and rotating wings \cite{eldredge_leading-edge_2019}.

The aerodynamic performance of the flapping membranes at different hovering frequencies ($f$) or membrane material properties ($E h$) can be described solely as a function of the aeroelastic number and the angle of attack amplitude $\hat{\alpha}$ for a trapezoidal pitching motion \cite{gehrke_aeroelastic_2022}.
We express the aerodynamic performance in terms of the stroke average lift coefficient \kindex{\bar{C}}{L} and the ratio between this stroke average lift coefficient and the stroke average power coefficient \kindex{\bar{C}}{P}, which is a measure of the stroke average efficiency $\bar{\eta}$. 

\section*{Impact of aeroelasticity on the vortex evolution}
\label{sec:vortex_evolution}
\begin{figure*}[tb]
	\centering
	\includegraphics[width=\textwidth]{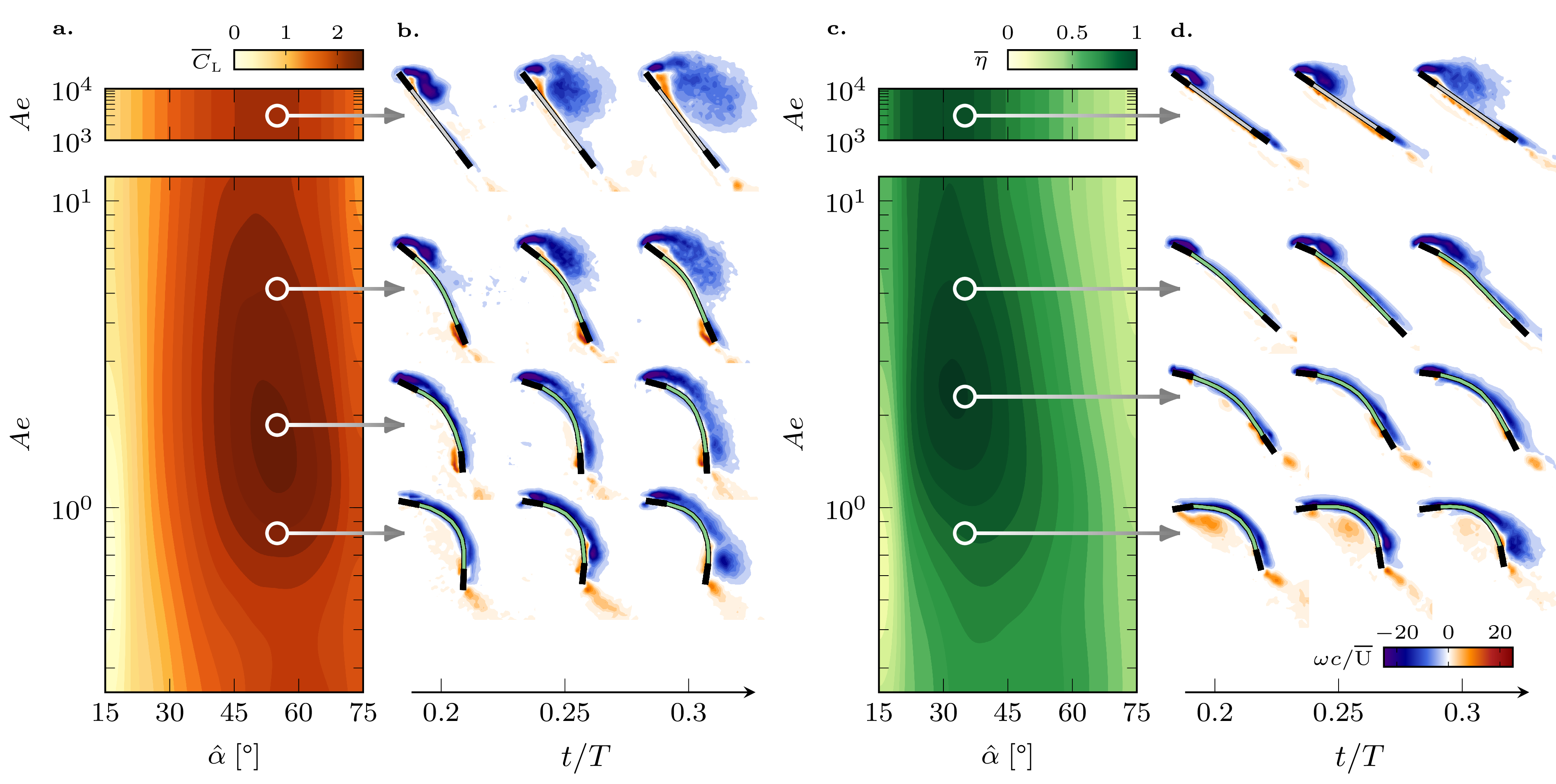}
	\caption{
		a.~Stroke-average lift coefficient \kindex{\bar{C}}{L}, and c.~power economy $\hat{\eta}$ contours as a function of the angle of attack amplitude $\hat{\alpha}$ and the aeroelastic number \Ae.
		b. and d.~Instantaneous vorticity fields and deformed camber lines are presented at $t/T = 0.20$, $0.25$, and $0.30$, for a fully rigid wing (top row) and three values of the aeroelastic numbers \Ae (indicated by the markers in the contour plots) at two angle of attack amplitudes: b.~$\hat{\alpha} = \ang{55}$ and d.~$\hat{\alpha} = \ang{35}$.
		The cycle-average lift coefficient is defined as $\kindex{\bar{C}}{L} = \bar{L} / (0.5 \rho \bar{U}^2 c R)$, and the power efficiency as the ratio $\bar{\eta} = \kindex{\bar{C}}{L} / \kindex{\bar{C}}{P}$, with the power coefficient $\kindex{\bar{C}}{P} = \bar{P} / (0.5 \rho \bar{U}^3 c R)$.
	}
	\label{fig:overviewPIVetaCl}
\end{figure*}%
The measured values of the stroke average lift coefficient and energy efficiency of our flapping membrane wings are presented in \cref{fig:overviewPIVetaCl} for a range of angle of attack amplitudes and aeroelastic numbers.
The global optima for the lift and the efficiency are found at different of aeroelastic conditions, expressed in terms of the aeroelastic number \Ae and the angle of attack amplitude $\hat{\alpha}$.
A maximum lift stroke average lift coefficient of $\kindex{\bar{C}}{L} = \num{2.43}$ is reached for $\Ae = \num{1.70}$ and $\hat{\alpha} = \ang{55}$ (\cref{fig:overviewPIVetaCl}a).
The highest stroke average power efficiency of $\bar{\eta} = \num{0.86}$ is reached at a lower angle of attack amplitude $\hat{\alpha} = \ang{33}$ and for slightly stiffer wings with a higher aeroelastic number $\Ae = \num{2.44}$ (\cref{fig:overviewPIVetaCl}c).

To understand why we obtain the best performances at these aeroelastic conditions, we conducted membrane deformation measurements and velocity field measurements using particle image velocimetry (PIV).  
The PIV data was obtained in a vertical cross-sectional plane at the radial location corresponding to the second moment of area $\kindex{R}{2}$ for the angle of attack amplitudes $\hat{\alpha} = \ang{55}$, where we found the highest stroke average lift, and $\hat{\alpha} = \ang{35}$, where we found the optimal efficiency.
Past experimental and numerical studies show that observations of the flow field at $\kindex{R}{2}$ provide a representative insight on of the vortex topology on flapping and revolving wings \cite{chen_vorticity_2023}.
Instantaneous vorticity fields and camber line deformations are presented here for $t/T = 0.20$, $0.25$, and $0.30$, with $T$ the full flapping period covering a front- and a backstroke, for different aeroelastic numbers in \cref{fig:overviewPIVetaCl}b and \cref{fig:overviewPIVetaCl}d for $\hat{\alpha} = \ang{55}$ and $\hat{\alpha} = \ang{35}$, respectively. 

The top rows in \cref{fig:overviewPIVetaCl}b and d show the snapshots corresponding to the fully rigid wings which show similar vortex topologies as previous flapping wing studies using rigid rectangular wings \cite{ellington_leading-edge_1996,sane_aerodynamics_2003,krishna_flowfield_2018,eldredge_leading-edge_2019,gehrke_phenomenology_2021}.
Early in the stroke, vorticity accumulates directly behind the leading edge and forms a coherent leading edge vortex (\cref{fig:overviewPIVetaCl}b,d first row, $t/T = 0.2$). 
This vortex grows between $t/T = 0.2$ and $t/T = 0.3$ but remains attached to the leading edge through its feeding shear layer. 
For the rigid wing at $\hat{\alpha} = \ang{55}$, secondary opposite-sign vorticity emerges between the feeding shear layer and the wing's surface.
The emergence of the secondary vorticity is a precursor for the separation of the leading edge vortex \cite{Kissing.2020}. 
The secondary vorticity is absent for the rigid wing at $\hat{\alpha} = \ang{35}$, where the strength and size of the vortex grows slower and the leading edge vortex stays closer to the wing's surface (\cref{fig:overviewPIVetaCl},d - first row).
The more compact and close-bound leading edge vortex for $\hat{\alpha} = \ang{35}$ promotes energy efficient hovering flight for rigid flapping wings, but it comes at the cost of less enhanced lift production.

When we introduce membrane compliance by reducing the aeroelastic number, the wings deform to a positively cambered thin airfoil, and the local angles of the leading edges with respect to the incoming flow decrease (\cref{fig:overviewPIVetaCl},b,d - second row).
The lower leading edge angle reduces the shear rate over the leading edge and allows the shear layer to move further along the wing's surface before it rolls up into a leading edge vortex.
As a result, the leading edge vortex grows more gradually, has a more elliptical shape, and stays closer to the wing.
The positive membrane camber in combination with the increased concentration of vorticity closer to the wing leads to an overall increase in both lift production and power economy compared to the rigid wings.

The third rows in \cref{fig:overviewPIVetaCl},b,d correspond to the optimal aeroelastic conditions.
Here, the leading edge aligns with the horizontal flow direction induced by the wing's stroke motion.
This alignment reduces the shear at the leading edge further such that no coherent leading edge vortex forms throughout the flapping wing cycle at the location of the measurement plane.
The flow stays attached to the wing and a layer of dense vorticity covers the entire upper side of the wing.
The membrane wings produce now the highest lift coefficients at high angles of attack, and the most power-efficient hovering at lower angles of attack despite suppressing the leading edge vortex formation entirely.
These results suggest that a leading edge vortex is not always required to generate high lift in flapping wing flight at low Reynolds numbers, and highly cambered flexible foils can provide even better aerodynamic performance.

If the relative flexibility of the compliant wings is increased below the optimal values of the aeroelastic number, the membrane curvature increases rapidly early in the cycle.
The curvature or camber is so strong that the flow separates in the aft portion of the chord, instead of at the leading edge, creating a vortex behind the wing, close to the trailing edge (\cref{fig:overviewPIVetaCl}b,d - fourth row).
The wing experiences a loss in lift and hovering efficiency due to the unfavourable location of the vortex and associated low pressure region behind the wing, which increases the drag and does not aid in lifting the wing as the leading vortex does.

Furthermore, for the lower angle of attack amplitude $\hat{\alpha} = \ang{35}$, the curvature of the membrane is so strong that the local leading edge angle becomes negative, which leads to the formation of a vortex on the pressure side, below the wing (\cref{fig:overviewPIVetaCl}d, fourth row).
This low pressure region below the wing induces additional drag and pulls the wing down, reducing its lift and power efficiency.  

These findings show that great improvements in aerodynamic performance can be achieved with the use of compliant flapping wings.
If the compliance is well-tuned, the flow separation and the formation of the leading edge formation is suppressed, leading to an increase in the lift production and the power economy of the system.	
However, improper aeroelastic design can lead to the formation of vortices behind and even below the wing which cause drastic losses in lift and increased drag.
To properly select and adapt the aeroelastic conditions, it is crucial to identify observable indicators of the aeroelastic wing performance and estimate an optimal region of operation as a function of these indicators.

\section*{Membrane deformation and aeroelastic scaling}
\label{sec:membrane_deformation}

\begin{figure*}[bt]
\centering
\includegraphics[]{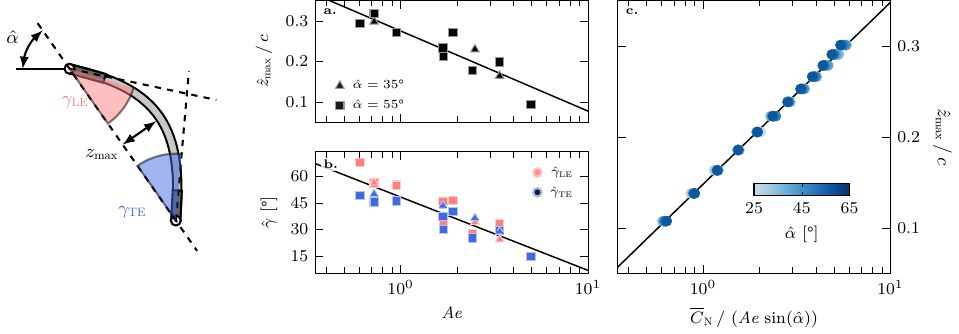}
\caption{
a.~Maximum wing camber $\kindex{\hat{z}}{max}$ during the stroke over aeroelastic number \Ae;
b.~maximal rotation angles $\kindex{\hat{\gamma}}{LE}$ at the leading edge and $\kindex{\hat{\gamma}}{TE}$ at the trailing edge during the stroke over aeroelastic number \Ae;
c.~Maximum wing camber $\kindex{\hat{z}}{max}$ over the scaled normal force coefficient $\kindex{C}{N}^*$ and normalized by the angle of attack.
The solid lines represent the general trend of the deformation data and are obtained by a least-square fit to the individual measurement data.
}
\label{fig:camberGammaVSAe}
\end{figure*}%

To identify observable indicators of the aeroelastic performance of our membrane wing, we take a detailed look at the wing deformation. 
The wing camber and the wing's relative orientation with respect to the flow are key parameters that characterize the aerodynamic performance \cite{gehrke_aeroelastic_2022}.
Our membrane wing deforms when it flaps through the fluid.
The fluid pushes the membrane up and creates a positive camber. 
The compliant wing's camber line has an approximately parabolic shape \cite{alon_tzezana_thrust_2019, gehrke_aeroelastic_2022}.
The wing camber varies along the chord length and in time, but is mainly uniform along the span despite the linear increase in the rotational velocity from root to tip.
This span-wise uniformity is largely attributed to the wing's specific design, which includes rigid reinforcements along the leading and trailing edges and thin wires at the root and tip to smoothly guide the trailing edge sliders.
The wing shape is thus described hereafter using a span-wise averaged representation.
Visual and quantitative evidence of the span-wise uniform membrane deformation is provided in the supporting information.

We determine the maximum camber \kindex{\hat{z}}{max} for a flapping motion characterized by a given aeroelastic number as the maximum membrane deflection relative to the undeformed state within a flapping cycle across the chord.
The maximum camber decreases with increasing aeroelastic number following in first approximation the same power law for both of the presented angle of attack amplitudes $\hat{\alpha} = \ang{35}$ and $\ang{55}$ (\cref{fig:camberGammaVSAe}a).
For a low relative wing stiffness, or a high wing loading, the membrane camber increases up to approximately \qty{0.3}{c}.

The shape of the membrane is further characterized by the leading and trailing edge rotation angles $\kindex{\hat{\gamma}}{LE}$ and $\kindex{\hat{\gamma}}{TE}$ (\cref{fig:camberGammaVSAe}b).
The leading and trailing edges of the wing platform are free to rotate.
As the wing reaches higher values of the maximum camber, the angles $\kindex{\hat{\gamma}}{LE}$ and $\kindex{\hat{\gamma}}{TE}$ increase proportionally.
The decrease in the rotation angles with the aeroelastic number can be approximated by a power law (\cref{fig:camberGammaVSAe}b).
The difference between the leading edge rotation angle and trailing edge rotation angle is minor ($<\ang{10}$) for most cases, and we consider the deformation symmetric with respect to the mid-chord (\cref{fig:camberGammaVSAe}b).
At the lowest tested aeroelastic numbers, the deformation is slightly asymmetric with the location of maximum camber closer to the leading edge and a larger difference between the leading and the trailing edge rotation angles.
We consider the shape asymmetry at low aeroelastic numbers a secondary effect that is omitted in the subsequent analysis.

The membrane deforms in response to the dynamic fluid pressure, and the shape of the membrane is characterized by $\kindex{\hat{z}}{max}$ and $\hat{\gamma}$, which decrease with \Ae following a power law (\cref{fig:camberGammaVSAe}a,b).
The shape of the membrane is governed by the balance between inertial forces (lift and drag) on the wing, and the tension of the membrane \cite{song_aeromechanics_2008, Ramananarivo.2011, waldman_camber_2017, gehrke_aeroelastic_2022}.
This balance can be expressed with a modified normal force coefficient $\kindex{C}{N}^*$:
\begin{equation}
\kindex{C}{N}^* = \frac{\kindex{C}{N}}{\Ae \sin (\hat{\alpha})} = \frac{N}{E h R \sin (\hat{\alpha})} \, ,
\label{eq:weber_number}
\end{equation}
with $N$ the normal force acting on the membrane.
Additional information about the deformation and aerodynamic force scaling relationship is detailed in the supporting information.

We find that all our data indeed collapse onto a single line when presenting the maximum camber $\kindex{\hat{z}}{max}$ versus $\kindex{\bar{C}}{N}^*$ in \cref{fig:camberGammaVSAe}c for different membrane thicknesses, various flapping frequencies, and a wide range of angles of attack amplitudes going from $\hat{\alpha} = \ang{25}$ to $\ang{65}$.
This scaling allows us to predict the membrane deformation based solely on the force measurements and vice versa, provided that we know a priori the flapping kinematics, characterised by $\hat{\alpha}$, the flapping frequency, and the membrane properties. 
The predictive scaling has interesting applications for the design and control of compliant membrane wings where either wing forces, membrane deformation, or membrane tension can be measured in flight and converted into the other quantities of interest.

We relate here the cycle average loading ($\kindex{\bar{C}}{N}$ and \Ae) with the deformation extrema ($\kindex{\hat{z}}{max}$ and $\hat{\gamma}$) and the extreme values govern the separation behaviour. 
Flow separation is inherently a threshold phenomenon, and occurs when critical angles of attack are exceeded, leading to significant performance changes.
Time-resolved force measurements reveal that staying below these angles prevents flow separation, resulting in higher peak forces, broader force profiles, and improved lift and energy efficiency during hovering of the flexible wings \cite{gehrke_aeroelastic_2022}.
\section*{Leading and trailing edge angles}
\label{sec:leading_trailing_angles}
\begin{figure*}
\centering
\includegraphics[]{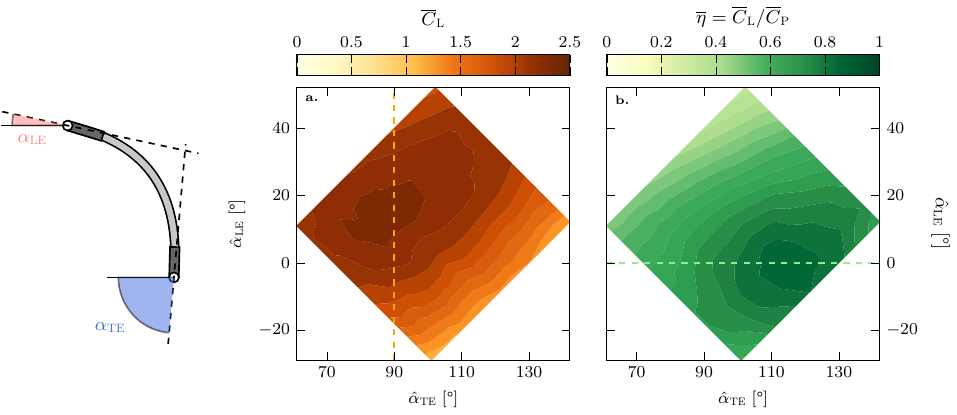}
\caption{
a.~Stroke-average lift coefficient $\kindex{\bar{C}}{L}$, and
b.~stroke-average power economy $\bar{\eta}$ for the membrane wing over leading $\kindex{\hat{\alpha}}{LE}$ and trailing edge angles $\kindex{\hat{\alpha}}{TE}$.
}
\label{fig:aeroPerf_alphaLETE}
\end{figure*}%
A coherent leading edge vortex is considered the key feature for sustained flapping wing flight.
The leading edge vortex is fed with vorticity by the separated shear layer at the leading edge.
The strength and orientation of the shear layer is determined by the relative flow velocity at the leading edge and the leading edge angle of attack $\kindex{\alpha}{LE}$ \cite{kriegseis_persistence_2013, eldredge_leading-edge_2019, gehrke_phenomenology_2021}.

We can now use the power law relations shown in \cref{fig:camberGammaVSAe} and assume symmetric deformation with respect to the mid-chord to calculate the leading and trailing edge angles relative to the flow induced by the wing's stroke motion for all our measured configurations based solely on the measured forces and the input kinematics and membrane properties.
This allows us to present the stroke-average lift coefficient $\kindex{\bar{C}}{L}$, and power economy $\bar{\eta}$, that were shown in \cref{fig:overviewPIVetaCl}a,c as a function of $\Ae$ and $\hat{\alpha}$, as a function of the cycle-minimum leading edge agle $\kindex{\hat{\alpha}}{LE}$ and cycle-maximum trailing edge angle $\kindex{\hat{\alpha}}{TE}$ in \cref{fig:aeroPerf_alphaLETE}. 

This allows us to link the location of the two distinct global optima for lift coefficient $\kindex{\bar{C}}{L}$, and power economy $\bar{\eta}$ to the physical angles between the leading and trailing edges and the flow.
The highest lift coefficient is reached when the maximum trailing edge angle is $\kindex{\hat{\alpha}}{TE} = \ang{90}$, corresponding to a trailing edge that points straight downwards (\cref{fig:aeroPerf_alphaLETE}a). 
The minimum leading edge angle corresponding to the highest lift coefficient is $\kindex{\hat{\alpha}}{LE} \approx \ang{10}$, which is close to the static stall angle of a flat plate.  
The maximum lift coefficient is thus achieved when the trailing edge points vertically downward to guide the flow around the wing downward to achieve the optimal vertical transfer of momentum.  
The leading edge angle is around the static stall limit of a flat plate to create the optimal suction peak while limiting massive leading edge separation. 

Conversely, the most power economic hovering is achieved when the leading edge is aligned with the incoming flow, $\kindex{\hat{\alpha}}{LE} = \ang{0}$. 
The maximum trailing edge angle for efficient hovering is $\kindex{\hat{\alpha}}{TE} \approx \ang{115}$, which points slightly more downstream than when the highest lift coefficient is achieved  (\cref{fig:aeroPerf_alphaLETE}b).
If the leading edge rotates past the optimum into negative angles ($\kindex{\hat{\alpha}}{LE} < \ang{0}$) or the trailing edge rotates past the vertical orientation ($\kindex{\hat{\alpha}}{LE} < \ang{90}$), the lift production and power efficiency decrease rapidly.
It is crucial to capture the instantaneous angle extrema thresholds for control purposes, as they not only influence peak loads but also trigger flow separation, resulting in significantly reduced overall performance.

\section*{Discussion}
\label{sec:vortex_phenomenology}
\begin{figure}[tb]
\centering
\includegraphics[]{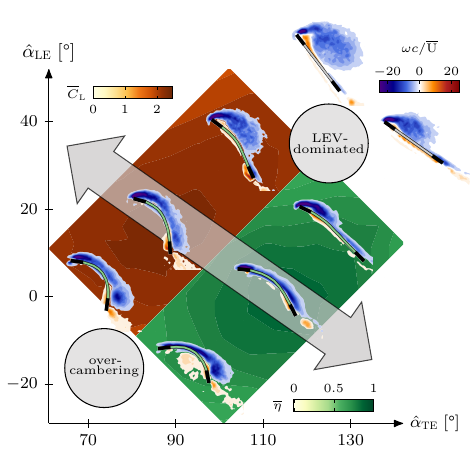}
\caption{
Vortex phenomenology and its relation to aerodynamic performance over the $\kindex{\hat{\alpha}}{LE}$ and $\kindex{\hat{\alpha}}{TE}$ space.
The positive and negative vorticity is depicted on the wings shortly after mid-stroke ($t/T = 0.30$).
The red and green contours correspond to the stroke-average lift coefficient $\kindex{\bar{C}}{L}$ and power economy $\bar{\eta}$ respectively.
In the top right are the rigid wings with large coherent leading edge vortices, common in hovering flapping wing flight.
In the middle band, indicated by the arrow, we have the region where the membrane flexibility aids to either optimize lift or efficiency by suppressing the leading edge vortex. 
This is the optimal region of operation for our flexible membrane wings.
If we continue further to the bottom left, the lift coefficient and efficiency drop due to over-cambering of the membrane. 
}
\label{fig:vortex_regions}
\end{figure}%

%
We summarize the emerging flow patterns, the associated wing deformation, and their relation to the aerodynamic performance in \cref{fig:vortex_regions}.
Two regimes of greater lift generating and higher power efficient hovering are indicated by the red ($\kindex{\bar{C}}{L}$) and green ($\bar{\eta} = \kindex{\bar{C}}{L} / \kindex{\bar{C}}{P}$) contour plots as a function of the leading and trailing edge angles of the membrane.
The different flow patterns are represented by the membrane shape and flow field snapshots taken just after mid-stroke ($t/T = 0.30$).
Videos showing the vortex evolution over the full cycle can be found in \supvid{1 - S4}.

The rigid wings in \cref{fig:vortex_regions} do not deform. 
The leading edge angle is equal to the angle of attack amplitude $\kindex{\hat{\alpha}}{LE} = \hat{\alpha}$ and the trailing edge angle is equal to $\kindex{\hat{\alpha}}{TE} = \ang{180} - \hat{\alpha}$.
The snapshots presented in \cref{fig:vortex_regions} have $\hat{\alpha}=\ang{35}$ or $\hat{\alpha}=\ang{55}$. 
Due to the strong shear at the leading edge, the flow separates early in the stroke cycle for these high angles of attack.
The separated shear layer rolls up into a large-scale coherent leading edge vortex.
During the unsteady stroke motion, the vortex grows until it covers most of the chord.
Then, the vortex lifts off of the wing and sheds into the wake.
At this time, the wing typically reverses its motion and a new leading edge vortex forms in the consecutive stroke.
This flow response is the typical behaviour observed for hovering insects and mechanical flapping wing devices using rigid wings.
The leading edge vortex growth creates a low pressure region on top of the wing which significantly augments the lift production on the wing compared to a static wing at the same angle of attack \cite{ellington_leading-edge_1996, sane_aerodynamics_2003, eldredge_leading-edge_2019}.

When flexibility is introduced, we move from the top right in \cref{fig:vortex_regions} towards the bottom left with increasing flexibility.
The slightly flexible wings produce more lift $\kindex{\bar{C}}{L}$ and the generation of the leading edge vortex requires less power than for the rigid wings, leading to a higher power economy $\bar{\eta}$ than the rigid wings.	
The membrane wings camber during flapping, and their leading and trailing edges rotate around their spanwise axes, reducing their local angle with respect to the flow.
A lower leading edge angle reduces the local shear at the leading edge, allowing the shear layer to stay closer to the wing's surface and the vorticity to be spread more evenly along the wing's chord.
This leads to leading edge vortices that are less concentrated for wings with low flexibility and the suppression of a distinct leading edge vortex altogether at optimal aeroelastic conditions.
At these optimal conditions, the vorticity no longer accumulates into a leading edge vortex but, instead, remains closer to the wing's surface and covers it from the leading to the trailing edge in the form of a bound shear layer.

The highest lift coefficients and the most energy efficient hovering are achieved when we observe a bound shear layer instead of a large scale leading edge vortex. 
A leading edge vortex increases the lift on the wing compared to a static wing at the same angle of attack as long as it is close to the wing. 
This additional lift is also referred to as circulatory lift or vortex lift \cite{Ford.2013, eldredge_leading-edge_2019}. 
The circulatory lift contribution is a function of the vorticity distributed above the wing, but this vorticity does not need to be present in the form of a coherent vortex. 
On the contrary, as we show here, high camber and a location of maximum camber close to the mid-chord lead to a uniform chordwise distribution of the vorticity close to the wing's surface, which results in a higher total lift production. 
The bound shear layer also prevents the emergence of a secondary vortex of opposite sign vorticity between the leading edge vortex and the wing's surface, which typically heralds the leading edge vortex pinch off and the decrease in vortex induced lift \cite{krishna_flowfield_2018,Kissing.2020}.
The delay in pinch-off allows for a more sustained generation of circulatory lift. 

When we go beyond the optimal region of operation, indicated by the arrow in \cref{fig:vortex_regions}, towards the bottom left, both stroke-average lift production and power economy decrease again.
Here, a low effective membrane stiffness leads to wing camber beyond \qty{20}{\percent} and large rotations of the leading and trailing edges that lead to two important changes in the flow topology.
Firstly, if the trailing edge points in the direction of flapping, a vortex is formed past mid-chord on the suction side of the wing due to flow separation aft of the maximum camber location.
Secondly, if the leading edge angle becomes negative, a vortex of opposite sign vorticity is formed below the wing.
In both cases, these vortices - contrary to the leading vortex on the rigid wings - are detrimental to the aerodynamic performance of the wing and reduce the stroke-average lift production and power economy.


Bats and other natural fliers using compliant membrane wings have been observed to favour lower angles of attack ($\alpha < \ang{40}$) for power efficient flight \cite{cheney_bats_2022, fan_power_2022}.
Our membrane wing has the most power economic lift generation at $\hat{\alpha} = \ang{35}$ which matches closely the values found in nature. 
Bats actively control the shape of their wings in flight by increasing the tension in their muscles to limit the curvature.
When their muscles are paralysed, and they can not adapt muscle tension anymore, they regulate the flight speed instead to control the wing camber \cite{cheney_bats_2022}.
By either regulating the wing flexibility $E h$ or the flight speed $U$, bats effectively regulate the aeroelastic number \Ae.
Our findings demonstrate how the modulation of the aeroelastic number affects the flow separation over membrane wings and controls the leading edge vortex formation.

Surprisingly, when our highly deforming flapping membrane wing performs best, the unsteady wing deformations suppress the formation of the leading-edge vortex.
Our findings thus suggest that a coherent leading-edge vortex is not necessary for sustained flapping wing flight if unsteady wing deformations create a highly cambered wing along which the vorticity can spread.
This behaviour is not commonly observed in natural flapping wing fliers. 
Flying insect rely on the formation and shedding of a strong, coherent leading-edge vortex \cite{Ellington.1984, Dickinson.1993}.
The flapping insect wings are thin membranes which are strengthened by a system of veins that limits the maximum amount of camber \cite{wang.2002,shyy_recent_2010}.
The wings of bats and other membrane wing fliers can reach higher values of camber than insect wings at hover \cite{vonbusse.2012, riskin.2010}.
In their seminal paper,  Muijres et al. \cite{muijres_leading_2014} suggest that bats are indeed able to control the development of leaving edge vortices and show examples of the flow around lesser long-nosed bat wings with leading edge vortices that are closer to the wing and stay attached throughout the downstroke.
At cruising speed, the formation of a prominent leading edge vortex is avoided which is suggested to increase efficiency.
This assumption is supported by our results for hovering kinematics and might also explain why small nectarivorous bats exhibit substantially lower metabolic compared to insects and birds of comparable size \cite{winter_energetic_1998, winter_energy_1998, voigt_energetic_1999}.

Based on the observed aeroelastic phenomena, active flow control schemes for human-engineered flapping membrane wing vehicles can be realized.
Optimal aeroelastic conditions can be maintained in flight by modulating either the wing's angle of attack $\alpha$, membrane stiffness $E h$, or flow velocity $U$.
The leading and trailing edge angles can serve as reliable indicators for the membrane shape and flow state and would allow the vehicle to maintain optimal aeroelastic conditions without additional sensory input.
A feedback control based on the instantaneous measurement of the leading and trailing edge angles would also be able to handle unsteady flow conditions.
Gusts, for example, can modify the relative velocity $U$ and the effective angle of attack $\alpha$, both of which are directly impacting the leading edge angle $\kindex{\hat{\alpha}}{LE}$.

\matmethods{%


\paragraph{Wing platform}
We cast rectangular membranes sheets from a silicone-based vinylpolysiloxane polymer (\textit{Zhermack Elite Double 32 shore A}) that has a density of $\rho = \SI{1160}{\kilogram\per\metre\cubed}$ and a Young’s modulus of $E = 1.22 \pm \SI{0.05}{\kilo\pascal}$ \cite{grandgeorge_mechanics_2021}.
The flexible membranes are then mounted on a rigid frame with leading- and trailing-edge brass bearings that allow the edges to rotate around their span-wise axes (\cref{fig:flexWingIntroduction}).
The assembled wings have a chord-length of $c = \SI{55}{\milli\meter}$ and a span of $R = \SI{150}{\milli\meter}$.
The membrane wings are covered with tracer markers which allows us to record the membrane shape over the entire flapping cycle using stereo photogrammetry.

\paragraph{Experiments}
All experiments were conducted in an octagonal water tank at a controlled temperature of $\SI{20}{\degreeCelsius}$ with a density of $\rho = \SI{998.23}{\kilogram\per\cubic\meter}$ and kinematic viscosity of $\nu = \SI{1.0023e-6}{\meter\squared\per\second}$.
The membrane wing model is mounted on a two-axis customised flapping wing test rig that mimics the stroke- and pitch-rotation of a single flapping wing in hover.
The aerodynamic forces are recorded with a 6-axis force/torque sensor connected at the wing root.
The stroke motion follows a sinusoidal motion with a peak-to-peak amplitude of $\hat{\phi} = \ang{90}$, and the pitch motion follows a trapezoidal profile with varying amplitudes from $\hat{\alpha} = \ang{15}$ to $\ang{75}$.
A video of the flapping wing in motion can be found in \supvid{1 - S4} and \cite{Gehrke.2021qjy}.

\paragraph{Flow field measurements}
To analyse the velocity field around the membrane wings, we conducted phase-locked PIV data at a fixed span-wise position corresponding to the radius of the second moment of area $\kindex{R}{2} \approx 0.55 R$ measured from the wing root.
A total of eight different flow field measurements were conducted varying the angle of attack amplitude $\hat{\alpha}$, and the aeroelastic number \Ae over the entire flapping wing cycle $T$.
Measurements were taken for the lift and power economy optimal aeroelastic conditions at $\hat{\alpha} = \ang{55}$ and $\ang{35}$.
At the same angle of attack amplitudes, additional measurements were taken for lower and higher than optimal \Ae, and the rigid reference case (\cref{fig:overviewPIVetaCl}).

\noindent Additional details about the experimental test platform, the wing kinematics, and measurements can be found in \cite{gehrke_phenomenology_2021, gehrke_aeroelastic_2022} or in the supporting information.

}
\showmatmethods{} 


\acknow{This work was supported by the Swiss National Science Foundation under grant number 200021\_175792.}

\showacknow{} 


\bibliography{bibliography.bib}

\pagebreak
\onecolumn
\section*{Supporting information}

\section*{Aeroelastic number estimation for hovering bats}
\subsection*{Morphological and kinematic parameters of hovering bat species}
To compare our results with aeroelastic wing deformation observed in hovering bat flight, we first estimate the aeroelastic number \textit{Ae} from morphological and kinematic observations of various biologic studies.
The aeroelastic number is defined as:
\begin{equation}
	\textit{Ae} = \frac{E h}{\frac{1}{2} \rho \overline{U}^2 \overline{c}} = \frac{E h}{2 \rho f^2 \hat{\phi}^2 c \kindex{R}{2}^2} \, ,
\end{equation}
where $\overline{c}$ is the average chord-length and $\kindex{R}{2}$ the radius of the second moment of area of the bat wings.
We assume standard air conditions at temperature of $\SI{20}{\degreeCelsius}$ and a density of $\rho = \SI{1.204}{\kilogram\per\cubic\meter}$.
Additionally, we need to determine the Young's modulus $E$, the wing thickness $h$, the wing stroke amplitude $\hat{\phi}$, the flapping frequency $f$, and the wing span $R$ for bats in hovering flight.
We were not able to find all important wing kinematic parameters for a single bat species.
Instead, we select three hovering bat species and collect and combine the relevant morphological and kinematic parameters to calculate \textit{Ae} from different sources.
The three species we select are small bats that use hovering extensively as part of their foraging and daily activities:
\begin{enumerate}
\item The \textbf{Pallas's long-tongued bat} (\textit{Glossophaga soricina}) hovers while feeding on nectar, similar to insects and hummingbirds.
It uses its long tongue to extract nectar from flowers~\cite{winter_operational_2003}.
\item The \textbf{lesser long-nosed bat} (\textit{Leptonycteris yerbabuenae}) hovers to feed on nectar from flowers, particularly those of cacti in desert regions across Mexico~\cite{cole_leptonycteris_2006}.
\item The \textbf{Geoffroy’s tailless bat} (\textit{Anoura geoffroyi}) hovers while feeding on nectar and pollen \cite{ortega_anoura_2008}.
\end{enumerate}
We collect the various morphological and kinematic parameters listed in \cref{tbl:bat_Ae_parameters} from the references \cite{swartz_mechanical_1996,norberg_ecological_1987,norberg_cost_1993,norberg_wing_2006}, and estimate the aeroelastic number for small hovering bats to be between $\kindex{\textit{Ae}}{min} = 0.10$ and $\kindex{\textit{Ae}}{max} = 4.33$.
If all parameter variations are averaged, we find a mean $\overline{\textit{Ae}} = 1.67$.
These \textit{Ae} estimations for hovering bats overlap with the range that we tested experimentally ($\textit{Ae} = 0.25$ to $12$) and the average value $\overline{\textit{Ae}}$ is close to the lift and efficiency optimal parameters ($\textit{Ae} = 1.7$ and $2.44$).

Note that for some parameters large biological variations are inherent, and for some parameters only limited data was available in the literature.
The authors could not find any references for the radius of the second moment of area $\kindex{R}{2}$ for the wings of the three species considered.
Following the procedure detailed in the next section, we estimate $\kindex{R}{2} = \SI{7.0}{\centi\meter}$ ($0.61R$) for the Pallas's long-tongued bat.
\begin{table}[h]
	\centering
	\caption{Morphological and kinematic parameters of hovering bats}
	\label{tbl:bat_Ae_parameters}
	\begin{tabular}{|l|l|l|l|}
		\hline
$E$   			& Young's modulus of the wing        	& $1 - \SI{10}{\mega\pascal}$   	& \cite{swartz_mechanical_1996} \\ \hline
$h$   			& Wing thickness                      	& $20 - \SI{100}{\micro\meter}$		& \cite{swartz_mechanical_1996} \\ \hline
$\hat{\phi}$ 	& Wing stroke amplitude (peak-to-peak)  & $\ang{120}$     					& \cite{norberg_cost_1993} \\ \hline
$f$   			& Wing stroke frequency               	& $\SI{12.7}{\hertz}$     			& \cite{norberg_wing_2006} \\ \hline
$\overline{c}$  & Mean wing chord length              	& $4.7 - \SI{7.4}{\centi\meter}$ 	& \cite{norberg_ecological_1987} \\ \hline
$2R$  			& Wing span (tip-to-tip)              	& $\SI{24}{\centi\meter}$       	& \cite{norberg_ecological_1987} \\ \hline
$\kindex{R}{2}$ & Radius of the second moment of area 	& $\SI{7.0}{\centi\meter}$      	& \ref{sec:bat_wing_R2} \\ \hline
	\end{tabular}
\end{table}
\subsection*{Radius of the second moment of area of a bat wing} \label{sec:bat_wing_R2}
\begin{figure}[h]
\centering
\includegraphics{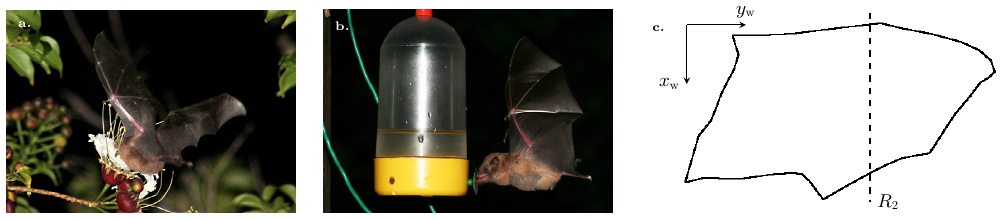}
\caption{
	Pallas's long-tongued bats in hover feeding on a.~the nectar from the flowers of a tree, and on b.~an artificial feeder filled with sugared water (adapted from \cite{sazima_two_2022}),
	c.~Extracted outline of the bat wing used for the calculation of the second moment of area and its radius. (Outline of a Pallas's long-tongued bat wing in hover, extracted from \cite{sazima_two_2022}).
}
\label{fig:batWingR2}
\end{figure}

We calculate the radius of the second moment of area $\kindex{R}{2}$ (also know as the radius of gyration) for a bat wing in hover as we could not explicitely find this information for the three considered species (Pallas's long-tongued bat, Lesser long-nosed bat, and Geoffroy's tailless bat) in literature.
We use the photograph of a Pallas's long-tongued bat in hovering flight (\cref{fig:batWingR2}a,b) and calculate $\kindex{R}{2}$ from the image.
First, we extract the outline of the wing and its pixel coordinates and extract the wing area~(\cref{fig:batWingR2}c).
Then, we calculate the radius of the second moment of area of the bat wing with respect to the stroke-axis according to:
\begin{equation}
	\kindex{R}{2} = \sqrt{\frac{\kindex{I}{xx} + \kindex{I}{yy}}{A}} \, ,
\end{equation}
where $\kindex{I}{xx}$ and $\kindex{I}{yy}$ are the second moments of area about the stroke- and pitch-axis respectively, and $A$ is the wing's area.
For this particular specimen we find $\kindex{R}{2} = \SI{7.0}{\centi\meter}$ (\cref{tbl:bat_Ae_parameters}).
\section*{Materials and methods}
\subsection*{Experimental setup}
\begin{figure}[h]
	\centering
	\includegraphics{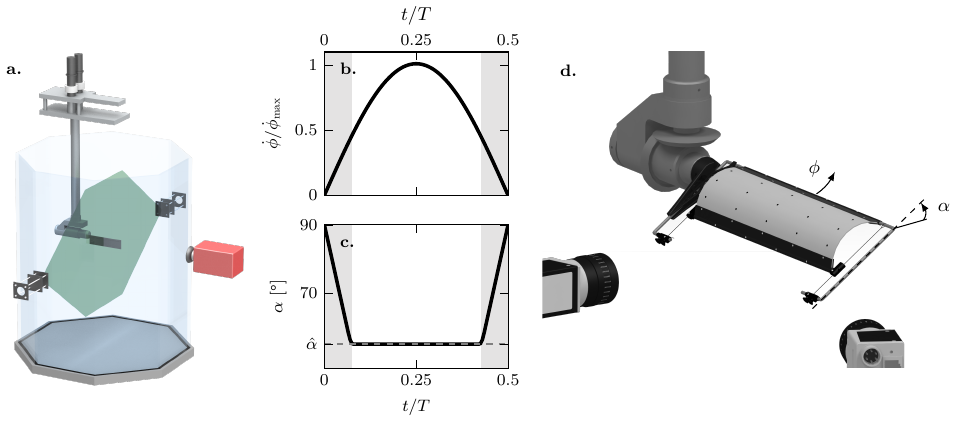}
	\caption{
		a.~Overview of the flapping wing robot insight the water tank with the PIV setup including two LEDs and a camera, b.~Flapping wing stroke velocity $\dot{\phi}$, and c.~angle of attack $\alpha$ over one half-cycle, d.~experimental setup sketch showing the stereo-deformation configuration (adapted from \cite{gehrke_aeroelastic_2022}).}
	\label{fig:kinematics}
\end{figure}%
The aerodynamic performance of different membrane wings and a rigid reference wing is evaluated using a robotic flapping wing device (\cref{fig:kinematics}a).
Experiments are conducted in a $\SI{0.75}{\meter}$ diameter octagonal tank filled with water at $\SI{20}{\degreeCelsius}$.
The tank is \SI{0.80}{\meter} high and the flapping wing mechanism is placed at mid-height. 
The mechanism itself is controlled by two servo motors (\textit{Maxon RE35, Switzerland}: $\SI{90}{\watt}$ power and $\SI{100}{\newton\milli\meter}$ torque), that guide the stroke and pitch axes.
The system executes complex, time-varying kinematics with a maximum error of $< \ang{0.1}$ between the control signal and motor response, ensuring high repeatability and robustness.

The wings consists of flexible membranes that are mounted on a rigid frame with leading- and trailing-edge brass bearings that allow the edges to rotate around their span-wise axes.
The assembled wings have a chord-length of $c = \SI{55}{\milli\meter}$ and a span of $R = \SI{150}{\milli\meter}$.
The stroke motion follows a sinusoidal motion with a peak-to-peak amplitude of $\hat{\phi} = \ang{90}$, and the pitch motion follows a trapezoidal profile with varying amplitudes from $\hat{\alpha} = \ang{15}$ to $\ang{75}$.
A minimum tip clearance of \SI{3.5}{c} is found for the tip-to-tip stroke-amplitude of $\hat{\phi} = \ang{90}$ which is deemed sufficient to avoid wall effects in flapping wing experiments \cite{manar.2014,krishna_flowfield_2018}. 
Initial force measurements at different distances from the wall were conducted to verify that for the selected stroke amplitude and frequencies no changes in the stroke-average forces are identified for wing tip to wall distances $> 3c$.

Aerodynamic loads are measured with a six-axis force-torque transducer (\textit{Nano17 IP68, ATI Industrial Automation, USA}) mounted at the wing root, with force and torque resolutions of $\SI{3.13}{\milli\newton}$ and $\SI{0.0156}{\newton\milli\meter}$, respectively.
The typical measured values vary from \SI{2}{x} the resolution for the slowest flapping wings to \SI{100}{x} the resolution for the faster flapping wings. 
Force signals are recorded at $\SI{1000}{\hertz}$ using a data acquisition module (\textit{NI-9220, National Instruments, USA}).
Instantaneous lift ($L$), drag ($D$), and pitch torque ($\kindex{T}{p}$) are obtained directly from the transducer.
Lift is defined as the upward force component perpendicular to the stroke plane, while drag is the force component in the stroke plane, positive when acting in the direction of the wing's velocity.
Aerodynamic power ($P$) is the sum of pitch power ($\kindex{P}{p}$) and stroke power ($\kindex{P}{s}$), calculated as $\kindex{P}{p} = -\kindex{T}{p} \dot{\alpha}$ and $\kindex{P}{s} = \kindex{T}{s} \dot{\phi}$.
The stroke torque ($\kindex{T}{s}$) is derived from the drag force across the span, $\kindex{T}{s} = \int_0^R D(r) r dr$, simplified to $\kindex{T}{s} = D \kindex{R}{d}$ with $\kindex{R}{d} = \frac{3}{4}\frac{(\kindex{R}{0} + R)^4 - \kindex{R}{0}^4}{(\kindex{R}{0} + R)^3 - \kindex{R}{0}^3}$.
Here, $\kindex{R}{0}$ is the distance from the stroke-rotation axis to the wing root also know as the root cutout and is equal to $\kindex{R}{0} = \SI{0.05}{\meter}$.

Forces and torques are recorded over 16 cycles.
We discard the first five cycles to remove transient effects.
The remaining 11 cycles are phase-averaged to obtain temporal evolutions and mean aerodynamic coefficients within the flapping period.
\subsection*{Kinematic parameters}
The kinematics are defined by the stroke axis $\phi$ which is perpendicular to the gravity or lift direction, and the wing's angle of attack alpha $\alpha$ (\cref{fig:kinematics}b-d).
The stroke angle $\phi$ of the flapping wing motion follows a sinusoidal motion with a peak-to-peak amplitude of $\hat{\phi} = \ang{90}$ (\cref{fig:kinematics}b).
The flapping frequency varies from $f = \SI{0.125}{\hertz}$ to $\SI{0.4}{\hertz}$ and defines the stroke average velocity $\overline{U} = \dot{\phi} r = 2 \hat{\phi} f r$ that the wing experiences at every span-wise location $r$.
The angle of attack of the wing $\alpha$ follows a trapezoidal profile and is varied amplitudes from $\hat{\alpha} = \ang{15}$ to $\ang{75}$ between experiments (\cref{fig:kinematics}c).
The reduced frequency $k$ for all experiments is kept constant and equal to $k = \pi c / (2 \hat{\phi} \kindex{R}{2}) = 0.42$ \cite{gehrke_aeroelastic_2022}.
\subsection*{Motion tracking}
The membrane shape is tracked over the entire cycle using stereo photogrammetry to quantify its impact on aerodynamic forces.
Deformation and load measurements are conducted simultaneously under specific experimental conditions for direct comparison.
Two CCD cameras (\textit{pco.pixelfly usb, ILA\_5150 GmbH/PCO AG, Germany}) at a $\ang{45}$ stereo angle, equipped with $\SI{12}{\milli\meter}$ focal length lenses, are used for marker tracking on the membrane wing throughout the flapping cycle (\cref{fig:kinematics}d).
With a camera resolution of $\SI{1392}{\px} \times \SI{1040}{\px}$, the spatial resolution ranges from $\SI{0.13}{\milli\meter\per\px}$ to $\SI{0.23}{\milli\meter\per\px}$, depending on the wing's distance from the cameras.
We record stereo images over 20 cycles, discarding the first five to remove transient effects.
The remaining 15 cycles are phase-averaged onto one half-cycle.
Images are recorded at $\SI{6.17}{\hertz}$, uncorrelated with the flapping frequency, and sorted by relative phase-time to achieve an effective acquisition rate of $\SI{186}{\hertz}$ per half-cycle.
\subsection*{Flow field measurements}
We measure the flow field at two angle of attack amplitudes ($\hat{\alpha} = \ang{55}$ and $\ang{35}$), corresponding to optimal lift and optimal power economy.
The wing stiffness is varied using four different aeroelastic numbers \textit{Ae}, resulting in eight cases, including optimal, suboptimal, and rigid reference wings (corresponding to the markers in figure 2 of the manuscript). 
Particle image velocimetry (PIV) is conducted in a plane normal to the span-wise direction at the radius of the second moment of area $\kindex{R}{2}$ for a rectangular wing planform (\cref{fig:kinematics}a).
A \SI{4}{\milli\meter} thick light sheet is produced using two high-power LEDs (\textit{LED Pulsed System, ILA\_5150 GmbH, Germany}) and cylindrical lenses.
The illuminated plane is recorded by a sCMOS camera (\textit{ILA\_5150 GmbH/PCO AG, Germany}) with a resolution of $\SI{2560}{\px} \times \SI{2160}{\px}$ rotated at $\ang{90}$, covering a $\SI{107}{\milli\meter} \times \SI{128}{\milli\meter}$ field of view.
Phase-locked PIV is performed by synchronizing the LED and cameras to record image pairs at specific phase angles $\phi$.
To capture different phase positions throughout the stroke cycle, the initial stroke angle is shifted relative to the LED plane.
A total of 70 stroke angle positions are recorded and averaged over 68 consecutive flapping cycles, discarding the first five cycles to remove transient effects.
A multi-grid algorithm with a $\SI{48}{\px} \times \SI{48}{\px}$ interrogation window size and 50\% overlap is used to correlate the images and reconstruct the velocity flow field with a physical resolution of $\SI{1.02}{\milli\meter}$ or $0.0186 c$.

The uncertainty in the instantaneous PIV velocity measurements is given by $\kindex{\epsilon}{vel}=\kindex{\epsilon}{dx}/(\Delta t\, M)$, with \kindex{\epsilon}{x} the single displacement error in pixel, $\Delta t$ the time interval between the snapshots correlated by the PIV algorithm, and $M$ the magnification factor \citep{raffel.2018}. 
As the stroke velocity of the flapping wing varies sinusoidally (\cref{fig:kinematics}b), we adapted the time interval between the PIV image pairs as a function of the phase angle to ensure sufficient particle displacement at any phase of the flapping motion. 
The single displacement error \kindex{\epsilon}{x} is the sum of the random error or measurement uncertainty \kindex{\epsilon}{rms} and the bias error \kindex{\epsilon}{bias}. 
The random error is estimated smaller than \SI{0.06}{px} in observation areas away from the wing and was found to increase up to \SI{0.25}{px} in observation areas close the wing's surface where strong velocity gradients are present. 
These estimation were conducted for selected representative snapshots based on the method proposed by Wieneke~\cite{wieneke.2015}.
The bias error is strongly affected by peak-locking, i.e. the tendency of displacements to be biased towards integer pixel values.
Histograms of subpixel displacement showed that peak locking was successfully avoided and the remaining bias error was assumed to be significantly less than the random error, i.e. $\kindex{\epsilon}{bias}<\SI{0.05}{px}$. 
By combining these estimates, we find a resulting relative PIV uncertainty $\kindex{\epsilon}{vel}/\overline{U}<0.03$. 

Supporting movies show the vorticity flow field over a full cycle for the rigid reference case and varies levels of effective wing stiffness (\url{doi.org/10.6084/m9.figshare.27146619.v1}).
%
\subsection*{Force and deformation scaling} \label{sec:force_def_scaling}
In the main manuscript, we relate the stroke-average aerodynamic forces (\kindex{C}{L}, \kindex{C}{D}) and power (\kindex{C}{P}) to the wing deformation  (\kindex{\hat{z}}{max}).
This allows us to predict the quantities of interest based on one measured quantity and provide tools for closed-loop control of flapping wing micro air vehicles.

We observe from the deformation data in figure 3a that the maximum camber \kindex{\hat{z}}{max} and rotation angles $\hat{\gamma}$ are insensitive to angle of attack amplitude $\hat{\alpha}$ changes within our experimental parameter range.
That allows us to predict the membrane geometry for any given set of experimental parameters $(f, h, \hat{\alpha})$ that yield $\textit{Ae} = E h / (0.5 \rho \overline{U}^2 c) = E h / (2 \rho f^2 \hat{\phi}^2 \kindex{R}{ref}^2 c)$.
Furthermore, we find that the aerodynamic forces in figure 2a have a strong coupling with $\hat{\alpha}$ and \textit{Ae}.
This motivates a force data scaling by using the membrane shape and the wings angle of attack, or its projected area with respect to the flow.

We start by defining the modified normal force coefficient similar to the Weber number introduced by \cite{waldman_camber_2017} and extended to flexible flapping wings in \cite{gehrke_aeroelastic_2022}:
\begin{equation}
	\kindex{C}{N}^* = \frac{\kindex{C}{N}}{\textit{Ae} \, \sin(\alpha)} = \frac{N}{\frac{1}{2}\rho R c U^2 \, \sin(\alpha)} \frac{\frac{1}{2}\rho c U^2}{E h} = \frac{N}{R E h \, \sin(\alpha)}.
	\label{eq:weber_number1}
\end{equation}%
Here, $c R \sin(\alpha)$ is the projected wing area exposed to the relative flow velocity $U$.
A similar expression can be found directly from a dimensional analysis normalizing the force coefficients of the flapping wing system where material properties ($E h$) are being used instead of the dynamic pressure ($0.5 \rho U^2 R c$).
Using the proposed normalization, only case dependent parameters remain, and we compare the cycle-average normal force $\kindex{\overline{C}}{N}^* = \overline{N} / \left(R E h \, \sin(\alpha)\right)$ with predicted camber maxima $\kindex{\hat{z}}{max} / c$ between all cases in figure 3c.
We find the relation between $\kindex{\overline{C}}{N}^*$ and $\kindex{\hat{z}}{max} / c$ can be expressed as a function of the form $f(x) = a \ln (x) + b$ with $a = 0.0869$ and $b = 0.148$.

We can use this expression to extrapolate our sparse deformation measurements over a wide range of parameters with the results from the force measurements ($\kindex{C}{N}$).
This gives access to the membrane shape parameter space (e.g. $\kindex{\hat{\gamma}}{LE}$ and $\kindex{\hat{\gamma}}{TE}$) that we use to present the aerodynamic performance $\kindex{\overline{C}}{L}$ and $\overline{\eta}$ in figures 4 and 5.
%
\subsection*{Long duration measurements of the lift force coefficient}
\label{sec:force_average_convergence}
\begin{figure}[tb]
\centering
\includegraphics[]{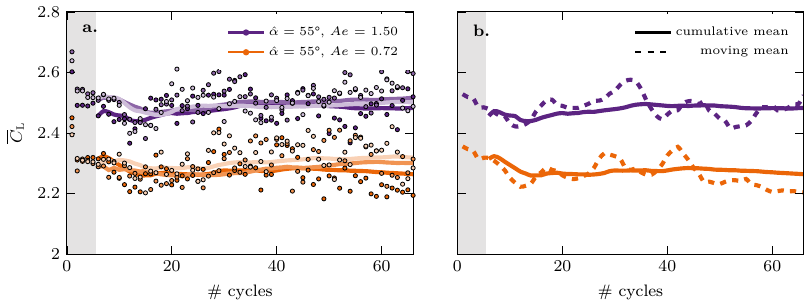}
\caption{
a. Stroke average lift coefficient recorded over 66 cycles for two different aeroelastic numbers $\Ae = 1.50$ and $0.72$ with $\hat{\alpha} = \ang{55}$.
The different colours refer to the different cases and the different shades indicate the different repetitions of the same parameter combination.
The markers indicate the cycle-average lift coefficient for the individual cycles and the solid lines represent the cumulative average lift starting from the sixth cycle.
b. Isolation of the results for only one of the three repetitions.
The solid lines represent again the cumulative average lift starting from the sixth cycle. 
The dashed lines represent the moving average over $5$ cycles.}
\label{fig:mCL_manyCycles}
\end{figure}%
To verify potential influences of the tank confinement on the average force data, we present in \cref{fig:mCL_manyCycles} the variation of the stroke average lift coefficient as a function of the flapping cycle for long duration measurements covering more than $60$ cycles.   
\Cref{fig:mCL_manyCycles}a summarises the data obtained for three repetitions of a long duration experiment covering $66$ flapping cycles for two different aeroelastic numbers $\Ae = 1.50$ and $0.72$ (corresponding to $f = \SI{0.225}{\hertz}$ and $\SI{0.325}{\hertz}$) at an angle of attack amplitude of $\hat{\alpha} = \ang{55}$.
The physical duration of these experiments is $5$ and $3.5$ minutes respectively.

The markers show the cycle-average lift coefficient for the individual cycles and the solid lines represent the cumulative average lift starting from the sixth cycle.
The grey area highlights the first five cycles which are removed from the averaging to exclude transient start-up effects.
\Cref{fig:mCL_manyCycles}b shows the results of only one out of the three repetitions for visual clarity.
The solid lines represent again the cumulative average lift starting from the sixth cycle. 
The dashed lines show a moving average over $5$ cycles to reveal low frequency variations that could potentially point to large-scale recirculation in the tank.
Even though we observe variations in the moving average lift, the cumulative average values do not change substantially after the first $16$ cycles.
The largest relative observed variation in the cumulative mean is found for the highest angle of attack amplitude case in \cref{fig:mCL_manyCycles}b where we notice a \SI{2}{\percent} variation between $\kindex{\overline{C}}{L,n=16} = 2.449$ and $\kindex{\overline{C}}{L,n=37} = 2.496$.
Overall, we do not observe any systematic drift in time or repeatable pattern that would indicate a recirculation in the tank that affects the results.
%
\subsection*{Force data resolution and wing inertia}
\label{sec:wing_inertia}
\begin{figure}[tb]
\centering
\includegraphics[]{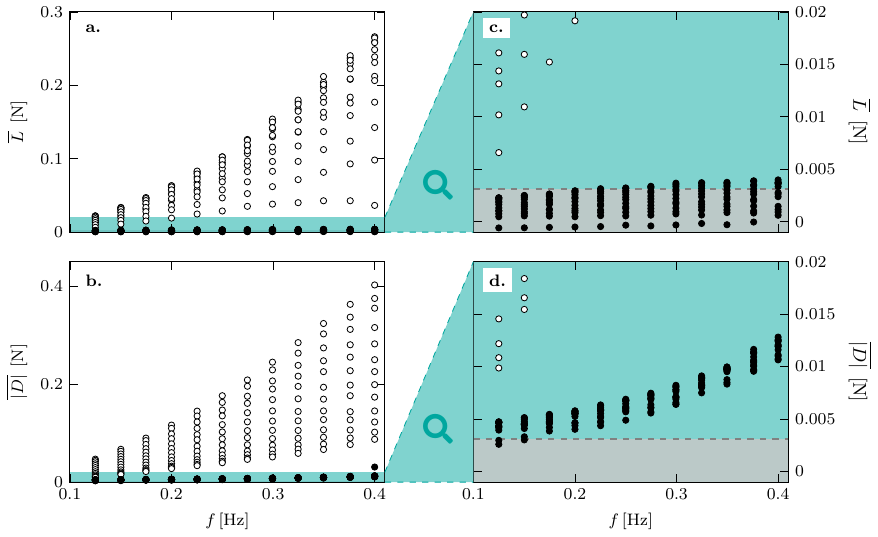}
\caption{
a. Dimensional average lift force $\overline{L}$ and b. average of the absolute drag force $\overline{\left| D \right|}$ for the membrane wing with thickness $t = \SI{1.4}{\milli\metre}$ as a function of the flapping frequency of the wing.
The white markers show the stroke-average forces in water and the black markers show the results in air.
c. and d. are zoomed in views of the data in a. and b., respectively. 
The grey area marks the resolution of the load cell. 
}
\label{fig:wing_inertia_quantification}
\end{figure}%

The ratio between inertial and aerodynamic forces for flapping wing depends strongly on the mass ratio between the wing and the fluid~\cite{gopalakrishnan_effect_2010}.
By conduction our experiments in water, we achieve low density ratios between the membrane wing and the fluid $\kindex{\rho}{m} / \kindex{\rho}{water} = 1.20$.
To estimate the effect of the inertial forces, we additionally conducted experiments in air for the same flapping kinematics.
In \cref{fig:wing_inertia_quantification}, we summarise the dimensional average lift force $\overline{L}$ and the average of the absolute drag force $\overline{\left| D \right|}$ for the heaviest membrane wing with thickness $t = \SI{1.4}{\milli\metre}$, which has the highest contribution of the inertial forces to the total measured force. 
The white markers show the stroke-average forces in water and the black markers correspond to the results of the measurements in air.
\Cref{fig:wing_inertia_quantification}a and b give an overview of the entire measurement set and \cref{fig:wing_inertia_quantification}c and d are zoomed in views of the coloured regions. 
The grey areas highlight the force transducer resolution of \SI{3.13}{\milli\newton} around zero.
Overall, The dimensional forces increase with increasing flapping frequency but the inertial forces recorded in air remain close to or stay even below the resolution of the load cell.
Therefore, we consider the inertial forces negligible for our set-up and their values have not been subtracted from the measured forces in water. 
%
%
\begin{figure}[t!]
\centering
\includegraphics[]{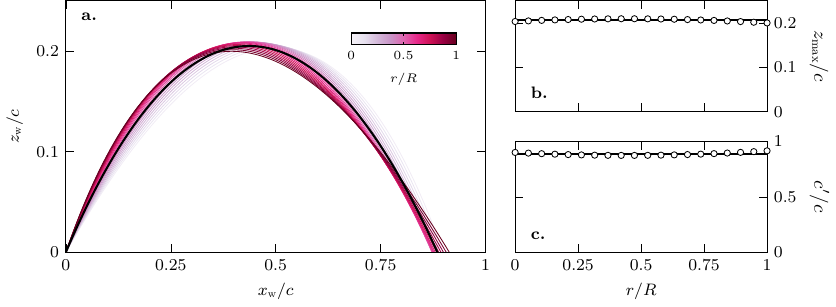}
\caption{
a.~Membrane wing camber-line in the wing's reference frame,
b.~maximum camber $\kindex{z}{max}$ as a function of the span-wise position ($r/R$), and
c.~effective chord-length $c^\prime$ as a function of the span-wise position ($r/R$).
The data shown here is recorded at mid-stroke ($t/T = 0.25$) for the optimal lift case ($\Ae = 1.70$ and $\hat{\alpha} = \ang{55}$).
The solid black line represents the average value in all three plots respectively.
}
\label{fig:spanwise_deformation}
\end{figure}%
\subsection*{Three-dimensional flow and membrane deformation} \label{sec:three_dimensional_deformation}
The flow around flapping wings is inherently three-dimensional, with the wing experiencing varying effective velocities across the span due to the stroke rotation.
Additionally, the stroke motion induces angular, centripetal, and Coriolis accelerations, which generate a span-wise velocity gradient from the wing root to the tip.
These factors influence the formation of the leading-edge vortex on flapping wings, and impact their aerodynamic performance \cite{lentink_rotational_2009}.

For flexible wings, such as those used in this study, the span-wise variation in wing loading is expected to influence the wing's deformation.
A rectangular wing typically experiences higher velocities and wing loading near the tip compared to the root, which feed the expectation of greater deformations towards the outboard sections.
However, our wing design employs two mechanisms to maintain uniform deformation along the span (\cref{fig:kinematics}d).
First, the leading and trailing edges are reinforced by rigid plates with a length of \SI{0.15}{c} to ensure consistent geometric leading and trailing edge angles along the span.
Second, thin wires are the root and tip guide the trailing edge sliders and limit spanwise variations in the effective chord-length $c^\prime$, defined as the distance between the leading end trailing edges of the deformed membrane wing.

These design features result in uniform deformation across the span as demonstrated in \cref{fig:spanwise_deformation} for the optimal lift case ($\Ae = 1.70$ and $\hat{\alpha} = \ang{55}$) at mid-stroke ($t/T = 0.25$), where the stroke velocity $\dot{\phi}$ is highest.
The camber-line $z/c$ as a function of chord-wise position $y/c$ shows minimal variation at different span-wise positions $r/R$ (\cref{fig:spanwise_deformation}a), with only a slight shift of maximum camber towards the trailing edge as $r/R$ increases. 
This trend is further reflected in the span-wise variation of maximum camber $\kindex{z}{max}/c$, with deviations of less than \SI{3.5}{\percent} from the mean (\cref{fig:spanwise_deformation}b).

As the wings deforms and camber increases, the trailing edge shifts towards the leading edge, preserving the arc length of the camber-line.
This deformation reduces the effective chord length $c^\prime$ decreasing the cross-sectional area exposed to the incoming flow compared to the undeformed chord length $c$.
The effective chord length $c^\prime$ also varies by less than \SI{3.5}{\percent} across the span $r/R$ (\cref{fig:spanwise_deformation}c), leading us to consider the membrane wing deformation approximately uniform.
In the manuscript, our analysis focuses on the span-averaged camber-line and its quantification. 
A supporting movie shows the flapping membrane wing in action (\url{doi.org/10.6084/m9.figshare.27146619.v1}). 
The movie illustrates the working principle of the membrane wing and its span-wise uniform passive deformation. 

\end{document}